\documentclass[aps,pra,twocolumn,nofootinbib,superscriptaddress,10pt]{revtex4-1}

\usepackage{natbib}
\usepackage{amsmath,amssymb,bm}
\usepackage{graphics}
\usepackage{collref} 

\newcommand{\im}[0]{{\rm i}}

\newcommand{\ket}[1]{\left|{#1}\right\rangle}
\newcommand{\bra}[1]{\left\langle{#1}\right|}
\newcommand{\braket}[2]{\left\langle{#1}|{#2}\right\rangle}

\begin{document}

\title{Optical orbital angular momentum under strong scintillation}

\author{Chemist M. Mabena}
\affiliation{CSIR National Laser Centre, P.O. Box 395, Pretoria 0001, South Africa}
\affiliation{School of Physics, University of the Witwatersrand, Johannesburg 2000, South Africa}

\author{Filippus S. Roux}
\email{froux@nmisa.org}
\affiliation{National Metrology Institute of South Africa, Meiring Naud{\'e} Road, Brummeria, Pretoria, South Africa}
\affiliation{School of Physics, University of the Witwatersrand, Johannesburg 2000, South Africa}

\begin{abstract}
The evolution of classical optical fields propagating through atmospheric turbulence is investigated under arbitrary conditions. We use the single-phase screen (SPS) method and the infinitesimal propagation equation (IPE), a multi-phase screen (MPS) method, to compute the optical power fractions retained in an input Laguerre-Gauss (LG) mode or transferred to higher order LG modes. Although they show the same trend while the scintillation is weak, the IPE and SPS predictions deviate when the strength of scintillation passes a certain threshold. These predictions are compared with numerical simulations of optical fields propagating through turbulence. The simulations are performed using an MPS model, based on the Kolmogorov theory of turbulence, for different turbulence conditions to allow comparison in both weak and strong scintillation. The numerical results agree well with the IPE results in all conditions, but deviate from the SPS results for strong scintillation.
\end{abstract}

\maketitle

\section{Introduction}

In paraxial optical fields, the two parts of the total angular momentum --- spin angular momentum (SAM) and orbital angular momentum (OAM) --- can be generated and controlled independently \cite{allenpo}. SAM, which is associated with the polarization state of the optical field, provides a two-dimensional configuration space, which is often used in quantum optics to represent a two-dimensional Hilbert space. OAM, on the other hand, is associated with the transverse spatial distribution of the optical field. It was found that certain modal bases such as the Laguerre-Gauss (LG) modes are OAM eigenstates carrying fixed quantized amounts of OAM \cite{allen}. Thus, it represents an infinite dimensional configuration space. The higher dimensionality of OAM makes it a potentially useful basis for applications in higher dimensional quantum information systems and multimodal communication systems \cite{leach,molina,pors2}. It is used for secure quantum communication in quantum cryptography and also for high-capacity free-space communication \cite{gibson,barreiro,mirhosseini}.

Free-space communication entails the propagation of optical fields through the atmosphere. Turbulence in the atmosphere induces random fluctuations in the index of refraction, which causes scintillation that distorts these optical fields \cite{scintbook}.

Various studies have been done to understand and clarify the effect of scintillation on the OAM of optical fields \cite{qturb4,qturb3,turboam2,pors,malik,oamturb,qkdturb, ipe,*iperr,lindb,notrunc,leonhard}. Most of the theoretical studies are based on a single-phase screen (SPS) approximation \cite{paterson}, which assumes that the effect of turbulence on a propagating optical field can be represented by a single random phase modulation. The SPS model is valid only under weak scintillation conditions \cite{turbsim}. Under strong scintillation conditions, the random phase modulation of the optical field is converted into intensity perturbation.

Strong scintillation conditions require a multi-phase screen (MPS) approach. The usefulness of such an MPS approach in numerical simulations has been shown by various authors for wave propagation in random media \cite{macaskill,flatte,buckley,rino}. It has also been used in classical studies of optical communications \cite{anguita}. Recently, an MPS based analytical approach has been employed to formulate the evolution of photonic quantum states propagating through turbulence in terms of an infinitesimal propagation equation (IPE) \cite{ipe,*iperr,lindb,notrunc}. Thanks to the MPS approach, the IPE is valid in both weak and strong scintillation conditions. Although it was proposed in the context of quantum optics, we show here that it also works for classical optical fields propagating through turbulence.

In this paper, we study the evolution of optical fields under arbitrary scintillation conditions. For this purpose, we consider an input optical field with an amplitude profile given by a certain OAM mode --- an LG mode with azimuthal index $\ell=1$ and radial index $p=0$. Then we investigate the effect of the scintillation on this mode by computing the overlap between the distorted LG mode and different LG modes to determine the fraction of optical power in the different modes. Both the IPE approach and the SPS approach are used to predict these power fractions as a function of propagation distance. We then compare these predictions against numerical simulations.

It is the first time that the IPE is subjected to a comparison with numerical simulation results. Although there are several numerical and experimental studies that tested the SPS model \cite{turboam2,pors,malik,oamturb,qutrit}, none so far have provided either experimental or numerical testing of the IPE. Here we use MPS numerical simulations that are based on the Kolmogorov theory of turbulence to determine whether the IPE predictions are better than the SPS predictions. We find that the numerical results agree well with the IPE predictions under all scintillation conditions. Since the IPE predictions deviate from the SPS predictions under strong scintillation conditions, the numerical results show that the IPE predictions are favored over those of the SPS approach.

The outline of the paper is as follows. In Sec.~\ref{agter}, we provide some theoretical background, as well as a discussion of the theoretical behavior of OAM modes in atmospheric turbulence. The details of the numerical simulation are discussed in Sec.~\ref{num}. The results are compared and discussed in Sec.~\ref{result} and in Sec.~\ref{concl} we end with some conclusions.

\section{\label{agter}Theoretical background}

\subsection{\label{scint}Atmospheric scintillation}

When light propagates through turbulence, the random phase modulations, induced by the fluctuating refractive index of the air, cause a distortion (scintillation) of the beam profile. Initially, the scintillation is weak, affecting only the phase of the optical beam. As the beam propagates further, the scintillation becomes progressively more severe and eventually also affects the intensity profile of the beam.

The statistical properties of a turbulent medium are often modelled in terms of the Kolmogorov theory \cite{scintbook}. The Kolmogorov phase structure function, as a function of the separation distance $x$, is given by
\begin{equation}
D(x) = 6.88 \left(\frac{x}{r_0}\right)^{5/3} .
\label{structf}
\end{equation}
It is expressed in terms of the Fried parameter \cite{fried}, which is defined as
\begin{equation}
r_0 = 0.185 \left(\frac{\lambda^2}{C_n^2 z}\right)^{3/5} ,
\label{fried}
\end{equation}
where $C_n^2$ is the refractive index structure constant, $\lambda$ is the wavelength and $z$ is the propagation distance. The turbulent medium can also be modelled by the Kolmogorov power spectral density \cite{scintbook}, which is given by
\begin{equation}
\Phi_n ({\bf k}) = 0.033 (2\pi)^3 C_n^2 |{\bf k}|^{-11/3} .
\label{klmgrv}
\end{equation}

Here, we'll represent the strength of the turbulence by a dimensionless quantity that naturally emerges from the analysis. It is given by \cite{turbsim}
\begin{equation}
\mathcal{K} = \frac{\pi^3 C_n^2 w_0^{11/3}}{\lambda^3} ,
\label{kdef}
\end{equation}
where $w_0$ is the Gaussian beam waist. Moreover, the propagation distance is represented as a dimensionless normalized propagation distance, by
\begin{equation}
t = \frac{z\lambda}{\pi w^2_0} .
\end{equation}

The strength of the scintillation is often quantified by the Rytov variance, which reads
\begin{equation}
\sigma_R^2 = 1.23 C_n^2 k_0^{7/6} z^{11/6} ,
\end{equation}
where $k_0=2\pi/\lambda$ is the wavenumber. It can also be expressed in terms of $t$ and $\mathcal{K}$:
\begin{equation}
\sigma_R^2 = 2.76 \mathcal{K} t^{11/6} .
\label{rytov}
\end{equation}
For weak scintillation, it is required that the Rytov variance is smaller than a constant of order 1 \cite{scintbook}.

\subsection{\label{IPE}Infinitesimal propagation equation}

The effect of a turbulent atmosphere on an optical beam can be determined with an SPS approximation under weak scintillation conditions. For strong turbulence however, an MPS approach is required. An approach that was recently introduced to model the evolution of photonic quantum states propagating through atmospheric turbulence is the infinitesimal propagation approach \cite{ipe,iperr,lindb,notrunc}, which leads to the IPE. It can however also be used for classical optical fields propagating through turbulence. Here, we briefly review the infinitesimal propagation approach.

The propagation of paraxial optical fields in turbulence is described by the stochastic parabolic equation
\begin{equation}
\partial_z f(\mathbf{r}) = -\frac{\im}{2 k_0}\nabla_\perp f(\mathbf{r})-\im k_0 \delta n(\mathbf{r}) f(\mathbf{r}) ,
\label{parB}
\end{equation}
where $\mathbf{r}$ is the three-dimensional position vector, $\delta n \left(\mathbf{r}\right)$ is the turbulence induced fluctuations of the refractive index of the atmosphere and $\nabla_\perp = \partial^2_x + \partial^2_y$. In the transverse Fourier domain, Eq.~(\ref{parB}) becomes
\begin{align}
\partial_z F(\mathbf{a},z) = & \im \pi \lambda |\mathbf{a}|^2 F(\mathbf{a},z) \nonumber \\
& - \im k_0 \int\delta N(\mathbf{a}-\mathbf{q},z) F(\mathbf{q},z)\ d^2q \label{Ftra},
\end{align}
where
\begin{align}
\begin{split}
F(\mathbf{a},z) & = \int f(\mathbf{x},z)\exp(\im 2\pi\mathbf{x}\cdot\mathbf{a})\ d^2x \\
\delta N(\mathbf{a},z) & = \int \delta n(\mathbf{x},z) \exp(\im 2\pi\mathbf{x}\cdot\mathbf{a})\ d^2x ,
\end{split}
\end{align}
with $\mathbf{x} = \{x,y\}$  and $\mathbf{a} = \{a_x,a_y\}$ being the two-dimensional transverse position coordinates and the transverse spatial frequency coordinates, respectively.

The density operator for a single-photon state can be expressed in the planewave basis, by
\begin{equation}
\hat{\rho}(z) = \int \ket{\mathbf{a}_1} \rho(\mathbf{a}_1,\mathbf{a}_2,z) \bra{\mathbf{a}_2}\ {d^2a_1}\ {d^2a_2} .
\end{equation}
For a pure state, the density `matrix' is a product
\begin{equation}
\rho(\mathbf{a}_1,\mathbf{a}_2,z) = F(\mathbf{a}_1,z) F^{*}(\mathbf{a}_2,z) ,
\end{equation}
where $F(\mathbf{a},z)=\braket{\mathbf{a}}{\psi(z)}$ is the angular spectrum or the Fourier domain wavefunction.

In the infinitesmal propagation approach, the evolution of the density matrix is described by a differential equation, which is called the IPE \cite{ipe, iperr, notrunc}
\begin{align}
\partial_z \rho(\mathbf{a}_1,\mathbf{a}_2,z) & = \im\pi\lambda\left(|\mathbf{a}_1|^2 - |\mathbf{a}_2|^2\right) \rho(\mathbf{a}_1,\mathbf{a}_2,z) \nonumber \\
 & + k_0^2 \int \Phi_0(\mathbf{q}) \left[\rho(\mathbf{a}_1-\mathbf{q},\mathbf{a}_2-\mathbf{q},z) \right. \nonumber \\
 & \left. - \rho(\mathbf{a}_1,\mathbf{a}_2,z)\right]\ {d^2q} ,
\label{FinalExpDens1}
\end{align}
where $\Phi_0(\mathbf{q})=\Phi_n(2\pi\mathbf{q},0)$. Under the quadratic structure function approximation \cite{leader}, the solution is \cite{notrunc}
\begin{align}
\rho & = \frac{\pi w_0^2}{4 \kappa t} \int \rho_0(\mathbf{a}_1-\mathbf{b},\mathbf{a}_2-\mathbf{b})\nonumber\\
 & \times \exp\bigg\{ -\pi^2 w^2_0 \bigg[ \frac{ \kappa t^3}{3} \left|\mathbf{a}_1-\mathbf{a}_2\right|^2 + \frac{|\mathbf{b}|^2}{4\kappa t} \nonumber \\
 & -\im t \left(|\mathbf{a}_1|^2-|\mathbf{a}_2|^2\right) - \im t(\mathbf{a}_1-\mathbf{a}_2)\cdot\mathbf{b} \bigg] \bigg\}\ d^2 b,
\label{1photonIPE}
\end{align}
where $\rho_0$ is the initial density matrix and $\kappa = 1.457 \mathcal{K}$.

One can use the single-photon expressions for classical optical beams. To consider scenarios where the photon state is entangled, one needs to generalize the single-photon expressions for two or more photons \cite{notrunc,kormed}.

\subsection{Generating function}

The input optical field is a single LG mode with radial index $p=0$. We are interested in the turbulence-induced coupling of this LG mode into the same or other LG modes with $p=0$. For the calculations, it is convenient to represent such LG modes by a generating function, given by
\begin{align}
\mathcal{G}(\mathbf{a};\mu) = & \mathcal{N} \pi w_0 \exp\left[ \im \pi w_0 \left( a_x \pm\im a_y \right) \mu \right. \nonumber\\
 & \left. - \pi w_0^2 \left(a_x^2 + a_y^2 \right)\left(1 - \im t \right) \right],
 \label{lGGen}
\end{align}
where $\mu$ is the generating parameter for the azimuthal index $\ell$, the sign in the first term in the exponent is given by the sign of the azimuthal index and the normalization constant is
\begin{equation}
\mathcal{N} = \left( \frac{2^{\ell+1} }{\pi |\ell|!} \right)^{1/2} .
\end{equation}
An LG mode is obtained from the generating function by
\begin{align}
{\cal M}_{\ell,0}\left(\mathbf{a}\right) = \left.\partial_{\mu}^{|\ell|}\mathcal{G}(\mathbf{a};\mu)\right|_{\mu=0} .
\label{operation}
\end{align}

\subsection{Overlap calculation}

For the calculation of the IPE solution, we substitute $\rho_0\rightarrow \mathcal{G}(\mathbf{a}_1;\mu_1) \mathcal{G}^*(\mathbf{a}_2;\mu_2)$ into Eq.~(\ref{1photonIPE}) and evaluate the integrals over $\mathbf{b}$. The result represents a generating function for the output density matrix.

For the case of an input LG mode with azimuthal index $\ell = 1$, the output density matrix is
\begin{align}
\rho = & \exp \left[ -\frac{\pi^2 w^2_0}{3 A_1} \left(A_4|\mathbf{a}_1|^2 - A_5 \mathbf{a}_1\cdot\mathbf{a}_2 + A^{*}_4|\mathbf{a}_2|^2 \right) \right] \nonumber\\
 & \times \left[\frac{\pi^4 w_0^4}{A_1^3}\left( A_2|\mathbf{a}_1|^2 + A_2^*|\mathbf{a}_2|^2 + A_3 \mathbf{a}_1\cdot\mathbf{a}_2 + A_6\right) \right. \nonumber\\
 & \left. + \im \frac{\pi^4 w_0^4}{A_1^2} (\mathbf{a}_2\times\mathbf{a}_1)\cdot \hat{z} \right] ,
 \label{abovIP}
\end{align}
where
\begin{align}
\begin{split}
A_1 & = 8\kappa t+1 \\
A_2 & = 2\im\kappa t^2 - 4\kappa^2t^4 - 16\kappa^2t^2 - 4\kappa t \\
A_3 & = 8\kappa^2t^4+32\kappa^2t^2+8\kappa t+1 \\
A_4 & = 8\kappa^2t^4+4\kappa t^3+12\kappa t-12\im\kappa t^2-3\im t+3 \\
A_5 & = 16\kappa^2t^4+8\kappa t^3+24\kappa t \\
A_6 & = \frac{32\kappa^2t^2 + 4 \kappa t}{\pi^2w_0^2} .
\end{split}
\end{align}
The expression in Eq.~(\ref{abovIP}) represents the state of the optical beam after the LG mode propagated through atmospheric turbulence.

The scintillation of the initial LG mode causes the optical field to contain a spectrum of LG modes that were not present in the initial optical field. In a free-space optical communication system where the different LG modes represent different channels, such a process would cause crosstalk among the different channels. One can quantify the amount of crosstalk with the aid of Eq.~(\ref{abovIP}).

To calculate the crosstalk, we evaluate the overlap between the output density matrix, given in Eq.~(\ref{abovIP}), and the generating function for LG modes, given in Eq.~(\ref{lGGen}). The overlap is expressed by
\begin{equation}
\eta = \int \rho(\mathbf{a}_1,\mathbf{a}_2,z) \mathcal{G}^*(\mathbf{a}_1;\mu_3)\mathcal{G}(\mathbf{a}_2;\mu_4)\ d^2a_1\ d^2a_2.
\end{equation}
The result is a generating function for the fraction of the input optical power that resides in the LG modes. To obtain the power fraction for a particular combination of input and output modes, one needs to apply the procedure given in Eq.~(\ref{operation}) four times, once for each of the four generating parameter. (Since we are dealing with the density matrix, both the input mode and the output mode appear twice in the expression.)

In Appendix \ref{appenda}, we provide the detailed expressions for the power fractions obtained in the LG modes with $\ell=1,2,3$ for an input LG mode with $\ell=1$. All these expressions depend only on the normalized propagation distance $t$ and the normalized turbulence strength $\mathcal{K}$.

\subsection{Weak scintillation limit}

The expression of the output state that is obtained from the IPE analysis, is valid under all scintillation conditions. Hence, one can reproduce the results under the SPS approximation from the IPE results in the limit of weak scintillation. The SPS results depend only on the dimensionless parameter $\mathcal{W}=w_0/r_0$, which can be expressed in terms of $\mathcal{K}$ and $t$ by
\begin{equation}
\mathcal{W} = 1.37 (\mathcal{K} t)^{3/5} .
\label{winkt}
\end{equation}
Using Eq.~(\ref{winkt}), one can replace $t$ in the IPE results in terms of $\mathcal{K}$ and $\mathcal{W}$. In the weak scintillation limit, it should then depend only on $\mathcal{W}$. To find out what happens to $\mathcal{K}$, one can express the Rytov variance in terms of $\mathcal{K}$ and $\mathcal{W}$ with the aid of Eq.~(\ref{winkt}). It then reads
\begin{equation}
\sigma_R^2 = 1.055 \frac{\mathcal{W}^{55/18}}{\mathcal{K}^{5/6}} .
\end{equation}
Since the Rytov variance represents the scintillation strength, it follows that the weak scintillation limit is obtain when $\mathcal{K}\rightarrow\infty$. When we apply this limit to the IPE results, they reproduce the SPS results. The SPS expressions for the power fractions in the weak scintillation limit that correspond to those obtained from the IPE calculation are provided in Appendix \ref{appenda}.

\subsection{Comparing weak and strong scintillation}

\begin{figure}[ht]
\includegraphics{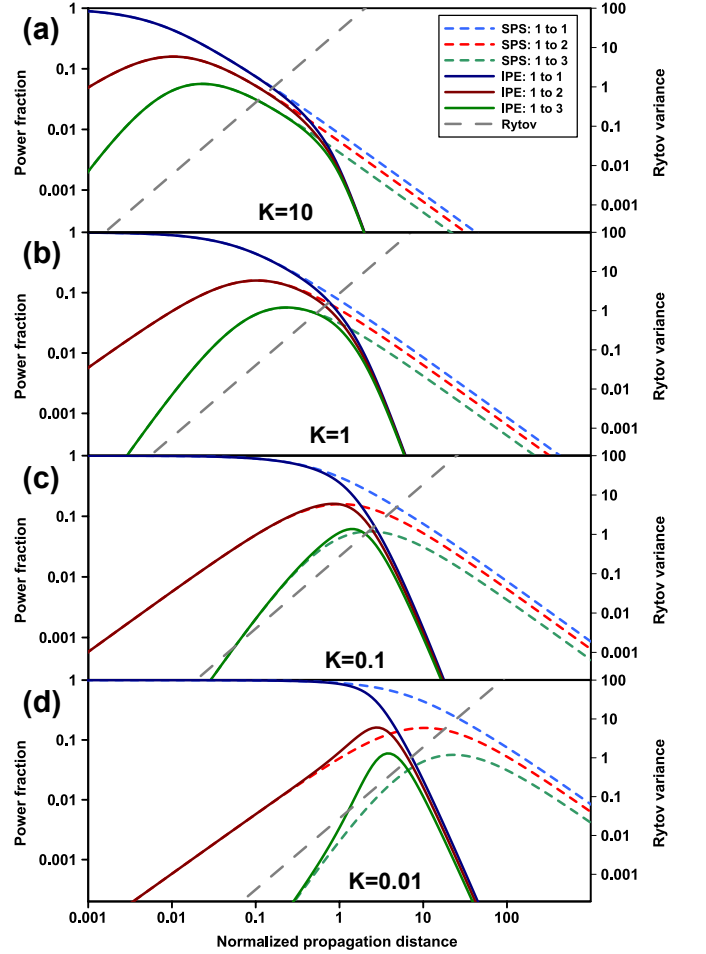}
\caption{Comparison of the predicted power fractions from the SPS approach (dashed lines) and the IPE approach (solid lines) as a function of $t$ for (a) $\mathcal{K}=10$, (b) $\mathcal{K}=1$, (c) $\mathcal{K}=0.1$ and (d) $\mathcal{K}=0.01$. Each graph shows the curves for the power fraction in output LG modes with $\ell=1,2,3$, respectively, for an input LG mode with $\ell=1$. The Rytov variance is provided in each graph (gray dashed line) with a separate vertical axis.}
\label{SPSIPEFig}
\end{figure}

In Fig.~\ref{SPSIPEFig}, we provide curves of the power fractions as a function of normalized propagation distance. The graphs for four different values of the normalized turbulence strength $\mathcal{K}=10,1,0.1,0.01$ are shown. Each of these graphs shows the curves for power fractions obtained in the output LG modes with $\ell=1,2,3$, for an input LG mode with $\ell=1$ as a function of $t$. The gray dashed line with the positive slope represents the Rytov variance $\sigma_R^2$ as it increases with propagation distance, labelled by the vertical axis on the left-hand side. All graphs in Fig.~\ref{SPSIPEFig} are shown as logarithmic plots.

There are various interesting observations that can be made from the graphs in Fig.~\ref{SPSIPEFig}. Focusing on the SPS curves (the dashed lines), we see that these curves are all governed by a single scale, which can be determined from the location of the region where the curves change their slopes. Away from these transition regions, the SPS curves are scale invariant, thanks to their power-law behavior (straight lines of a logarithmic plot). An expression for this scale can be obtained by noticing that when the SPS curves are plotted as a function of $\mathcal{W}$, the transition region is always located close to $\mathcal{W}\approx \sqrt{\ell}$. By solving this equation for the propagation distance, one finds that the distance scale is
\begin{equation}
z_{ws} \approx \frac{0.06 \lambda^2 \ell^{5/6}}{w_0^{5/3} C_n^2} .
\end{equation}
We'll refer to it as the {\em weak scintillation scale}. Previously, it was identified as the scale at which quantum entanglement decays to zero \cite{oamturb,leonhard,qkdturb}. However, such an observation was based on the fact that all those analyses were done for quantum entanglement within the SPS approximation. When we consider the situation for classical OAM coupling under strong scintillation conditions, we find that the situation is more complicated.

We also see from Fig.~\ref{SPSIPEFig} that by varying $\mathcal{K}$, one simply causes a horizontal shift of the SPS curves. The shapes of the SPS curves remain the same regardless of the value of $\mathcal{K}$. A shift on a logarithmic axis implies a scaling on a linear axis $t\rightarrow \alpha t$, for some scale factor $\alpha$. Hence, smaller values of $\mathcal{K}$ simply mean that the curves are stretched out over larger propagation distances.

The IPE curves in Fig.~\ref{SPSIPEFig} (solid lines) start out following the SPS curves, but then at a certain point, when the scintillation strength crosses some threshold, they start to deviate from the SPS curves. Beyond these points, the IPE curves become horizontally-squashed versions of the remainder of the SPS curves. A scaling on a logarithmic axis represents a change in the power of the independent variable $t\rightarrow t^{\gamma}$, with $\gamma>1$. We can see that those parts of the SPS curves that represent power laws again correspond to power laws for the IPE curves. Hence, we conclude that the squashing is done uniformly beyond the transition region. In other words, $\gamma$ does not depend on $t$ beyond the transition region.

The points where the IPE curves start to deviate from their SPS counterparts represent an additional scale that differs from the weak scintillation scale. We call the new scale the {\em deviation scale}. We see that when the turbulence is strong ($\mathcal{K}$ is large), the deviation scale is larger than the weak scintillation scale; it lies at a larger propagation distance. On the other hand, when the turbulence is weak, as indicated by a smaller value for $\mathcal{K}$, the deviation scale is found at shorter propagation distances than the weak scintillation scale. As a result, one finds (somewhat counterintuitively) that weak turbulence causes the IPE curves to deviate from the SPS curves at an earlier stage relative to the shape of the SPS curves. This behavior can be seen more clearly when the curves are plotted as a function of $\mathcal{W}$ instead of $t$. In such a case, all the SPS curves for a given overlap lie on top of each other regardless of the value of $\mathcal{K}$ and the IPE curve for the weakest turbulence (smallest $\mathcal{K}$) deviates quicker (at a smaller value of $\mathcal{W}$) than the others.

The notion that the deviation of the IPE curves from their SPS counterparts is an indication of the onset of strong scintillation does not always seem to agree with the value of the corresponding Rytov variance at these points. One can use the Rytov curves in Fig.~\ref{SPSIPEFig} to determine the value of the Rytov variance at the point where the IPE curves start to deviate from the SPS curves: draw an imaginary vertical line through the points on the three curves where they start to deviate. The crossing point of this vertical line and the Rytov curve then indicates at what value of the Rytov variance the deviation occurs. We find that when the turbulence is strong, the Rytov variance at the point of deviation is about equal to 1, which is what one would expect if the deviation is an indication of the onset of strong scintillation. However, for weak turbulence, the deviation seems to occur at a much smaller value of the Rytov variance. The value can be as much as an order of magnitude smaller. Therefore, it is questionable whether the deviation would indicate the onset of strong scintillation in such weak turbulence conditions. A better explanation may be that the deviation simply indicates the point where the SPS approximation breaks down, for whatever reason. The nature of the deviation scale is therefore a matter to be investigated further.

\section{\label{num}Numerical simulation}

To simulate the propagation of an optical field through a turbulent atmosphere, one needs to incorporate two different aspects of the process: refraction and diffraction, which are, respectively, represented by the two terms on the right-hand side of the stochastic parabolic equation in Eq.~(\ref{parB}). For this reason, we employ a split-step method \cite{mf1,mf2,anguita} to perform the numerical simulations. In one step, the optical field is modulated by a random phase screen, taking into account the refractive part of the process. In the subsequent step, the modulated field is propagated for a short distance through free-space, representing the diffractive part of the process. It is then followed by another modulation with a different random phase screen, then by another free-space propagation, and so forth. By repeating these two steps (the bi-step) several times, one can obtain the optical field after a significant distance of propagation through a turbulent atmosphere. As such, it is an implementation of a MPS approach. One such sequence of repeated bi-steps, consecutively applied to a particular input field, represents one single realization of a turbulent medium. In the numerical simulations, several realizations are performed for the same conditions and the same input field. It allows one to build up statistics from which the observables can be calculated as average values, together with standard errors.

If the bi-step is performed only once, the second step becomes superfluous, because it is compensated by the same free-space propagation of the modes with which it is overlapped. As a result, a single iteration of the bi-step is equivalent to the SPS approach. It thus follows that each bi-step in the MPS approach must obey the requirements for weak scintillation. Therefore, to implement the split-step process, one needs to divide the propagation distance $z$ into a series of slices $\Delta z$ in such a way that each slice represents a weakly scattering medium. Each phase screen is designed to represent the random phase modulation caused by the turbulent medium in one such slice.

Note that the phase modulations do not affect the intensity profile of the optical field directly. The scintillation of the intensity is brought about by the combined effect of the random phase modulations and the free-space propagations. It appears only after the beam has propagated some distance, which would involve several bi-steps.

Each random phase screen is computed as the inverse Fourier transform of a spectral representation of the random medium in term of filtering Gaussian noise. The spectral representation is composed of a 2D-array of normally distributed random complex numbers, with a zero mean and a unit variance, multiplied by the square root of the Kolmogorov power spectral density as an envelope function. It is given by \cite{mf1,knepp}
\begin{equation}
\theta = \left(2 \pi k_0^2\Delta z \right)^{1/2}\mathcal{F}^{-1}\left\{ \chi(\mathbf{a}) \left[ \frac{\Phi_0(\mathbf{a})}{\Delta^2_{k}} \right]^{{1}/{2}}\right\},
\label{PhScr}
 \end{equation}
where $\Delta z$ is the partitioned propagation distance between two phase screens, $\mathcal{F}^{-1}$ denotes the inverse Fourier transform,
$\Phi_0(\mathbf{a})$ is the Kolmogorov power spectral density, as defined below Eq.~(\ref{FinalExpDens1}), $\Delta_{k}$ is the grid size in the spatial frequency domain, and $\chi(\mathbf{a})$ is a normally distributed complex random function. The latter has zero mean and is $\delta$-correlated
\begin{equation}
\langle \chi(\mathbf{a}_1)\chi^{*}(\mathbf{a}_2)\rangle = \Delta_{k}^2 \delta(\mathbf{a}_1-\mathbf{a}_2) ,
\end{equation}
where $\langle\cdot\rangle$ denotes an ensemble average. The procedure in Eq.~(\ref{PhScr}) produces a random phase function that is complex-valued. As a result, each calculation produces two independent phase screens: the real and the imaginary parts of the resulting complex array.

An important issue with the above described method of generating the phase screens is that it does not consider the effect of large eddies; the discrete Fourier transform excludes the contributions of the lower spatial frequencies smaller than the grid size on the frequency domain. Due to the shape of the Kolmogorov power spectral density, these lower frequency components dominate. Their exclusion results in an inaccurate representation of the statistical nature of the scintillation process. To improve the accuracy, one can add more of the lower frequency components to the resulting phase function. The method of sub-harmonics \cite{dainty} produces an additional phase function that is added to the one generated by the Fourier calculation in Eq.~(\ref{PhScr}). The resulting phase screen becomes $\theta \rightarrow \theta+\theta_{SH}$, where
\begin{align}
\theta_{SH}(m\Delta_x, n\Delta_y) & = \sum_{r=1}^{N_s} \sum_{p,q=-1}^{1} \left( \mu_{p,q,r} + \im \nu_{p,q,r} \right) \nonumber \\
 & \times \exp\left[ \im 2\pi\left(p\Delta_r m\Delta_x + q\Delta_r n\Delta_y\right) \right] .
\end{align}
The variances of the randomly generated $\mu$ and $\nu$ are
\begin{equation}
\langle \mu_{p,q,r}^2\rangle = \langle \nu_{p,q,r}^2\rangle = \Delta_r^2 \Phi_{\theta}(p\Delta_r ,q\Delta_r),
\end{equation}
where $\Delta_r = \Delta_k/3^r$ and $N_s$ is the number of sub-harmonics.

\section{\label{result}Results}

\begin{figure}[ht]
\includegraphics{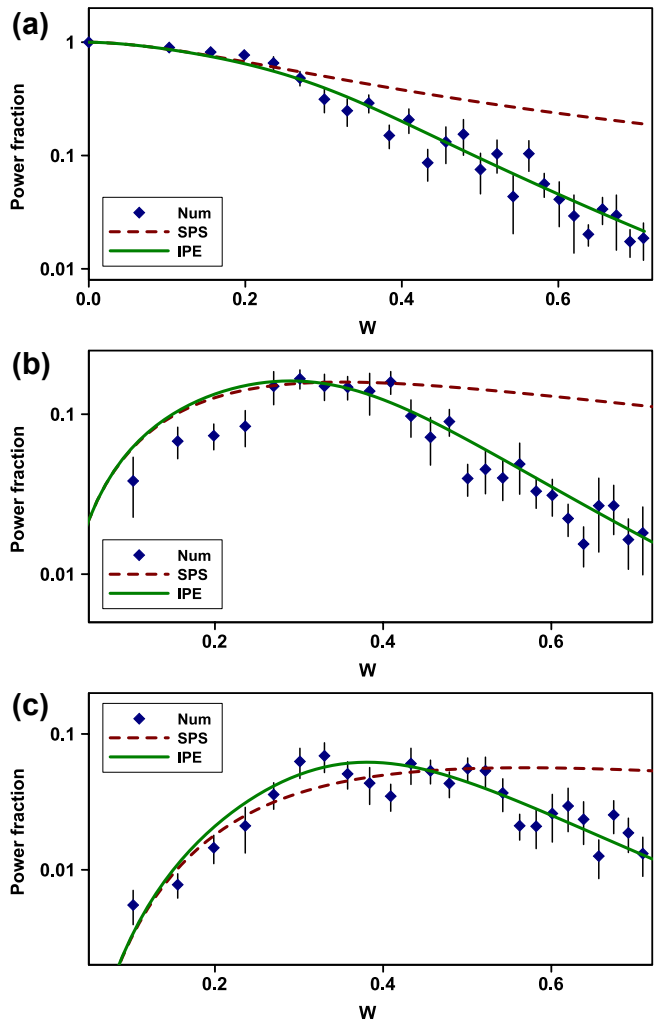}
\caption{Power fraction as a function of $\mathcal{W}$ for an input LG mode with $\ell=1$ propagating through weak turbulence with a normalized turbulence strength of $K=0.07$. The graphs show the fractions of optical power obtained in the LG modes with (a) $\ell=1$; (b) $\ell=2$; and (c) $\ell=3$, respectively. The markers represent the average power fraction and the error bars indicate the standard error (standard deviation of the mean). The solid lines (dashed lines) represent the theoretical IPE (SPS) predictions.}
\label{sim007}
\end{figure}

\begin{figure}[ht]
\includegraphics{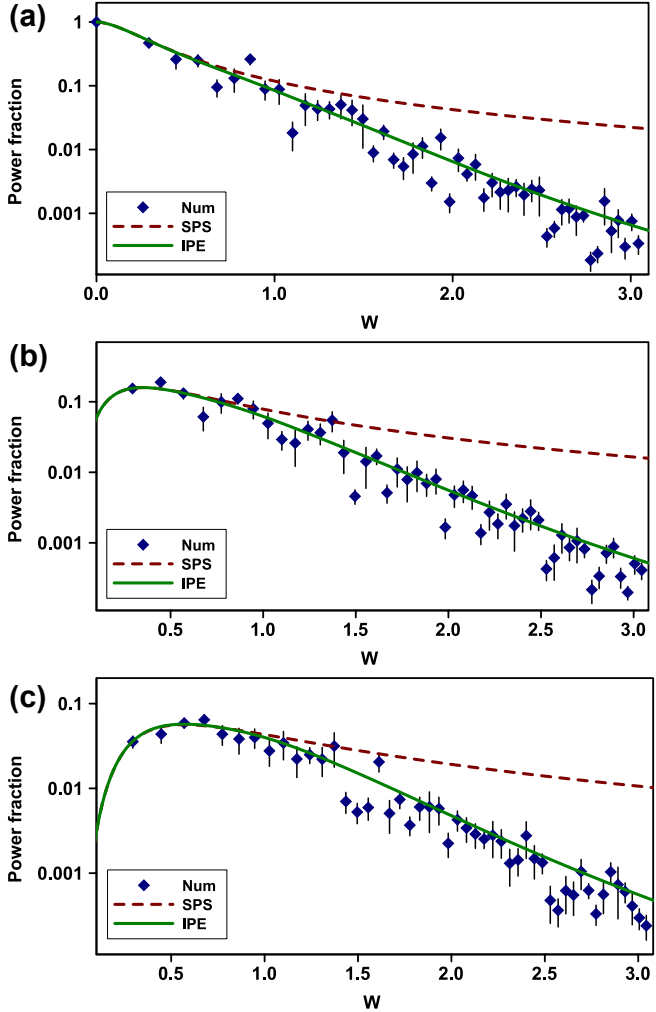}
\caption{Power fraction as a function of $\mathcal{W}$ for an input LG mode with $\ell=1$ propagating through stronger turbulence with a normalized turbulence strength of $K=0.7$. The graphs show the fractions of optical power obtained in the LG modes with (a) $\ell=1$; (b) $\ell=2$; and (c) $\ell=3$, respectively. The markers represent the average power fraction and the error bars indicate the standard error (standard deviation of the mean). The solid lines (dashed lines) represent the theoretical IPE (SPS) predictions.}
\label{sim07}
\end{figure}

We performed numerical simulations for the evolution of an input LG mode with $\ell=1$ under two different turbulence conditions: weak turbulence with $\mathcal{K}=0.07$ and stronger turbulence with $\mathcal{K}=0.7$. The resulting power fractions for $\mathcal{K}=0.07$ are shown in Fig.~\ref{sim007} and those for $\mathcal{K}=0.7$ are shown in Fig.~\ref{sim07}. All curves are plotted as functions of $\mathcal{W}$. In each of the two cases, we plot separate graphs for the output LG modes with $\ell=1,2,3$, respectively. In each graph, we show the IPE predictions as solid lines and the SPS predictions as dashed lines. The numerical simulation results are shown by discrete markers. They represent the average power fractions, obtained from a thousand runs for each of the two turbulence conditions. The error bars represent the standard error (standard deviation of the mean).

The range of values of $\mathcal{W}$ are chosen to show a region where the IPE curves and the SPS curves deviate from each other. In all the graphs in Figs.~\ref{sim007} and \ref{sim07}, the numerical results follow the IPE curves. This agreement indicates that the IPE predictions are more reliable than the SPS predictions under strong scintillation conditions.

The first graphs for both conditions Figs.~\ref{sim007}(a) and \ref{sim07}(a) represent the fraction of optical power that remains in the input mode (LG mode with $\ell=1$) during propagation through turbulence. Although the initial trend is the same for both the IPE and the SPS curves, a point is reached where these two curves start to deviate from one another. The IPE curves start to decay more quickly than the SPS curves. In both cases, the numerical results follow the IPE predictions more closely. However, in Fig.~\ref{sim07}(a), it seems that the IPE prediction is slightly higher than the trend of the numerical results. The reason may be found in the fact that the IPE prediction incorporates the quadratic structure function approximation for the sake of tractability, whereas the numerical simulations do not incorporate this approximation.

The other two graphs for both conditions Figs.~\ref{sim007}(b), \ref{sim007}(c), \ref{sim07}(b) and \ref{sim07}(c) represent the fraction of optical power that is transferred from the input mode (LG mode with $\ell=1$) to higher order modes (LG modes with $\ell=2$ and $\ell=3$). Again, the initial trends in all these graphs are the same for both the IPE and the SPS curves. They show how the power fraction increases from zero as more and more optical power is transferred from the input mode into these higher order modes. Then at some point, the IPE curves start to deviate from the SPS curves. In Figs.~\ref{sim07}(b) and \ref{sim07}(c), the deviation starts after the curves have passed their peaks. Optical power is now lost to other modes of even higher order. The IPE curves in Figs.~\ref{sim07}(b) and \ref{sim07}(c) drop below the SPS curves as they start to experience an accelerated decay. In Figs.~\ref{sim007}(b) and \ref{sim007}(c), the deviations between the IPE and the SPS curves start before the curves reach their peaks. As a result, the IPE curves at first rise above the SPS curves and cross the SPS curves to drop below them in an accelerated decay. In all these graphs, the numerical results follow the IPE predictions. However, due to the fluctuations in the numerical data, one cannot make a clear distinction between the two predictions when they are still close together. It is only when the IPE curves have dropped significantly below the SPS curves that one can see that the numerical data follow the IPE trend. In Figs.~\ref{sim07}(b) and \ref{sim07}(c), we again see a slightly lower trend in the numerical data than what the IPE trend seems to predict, which is again believed to be due to the difference in the way the turbulence is modeled (quadratic structure function approximation in the IPE vs. Kolmogorov power spectral density in the numerical simulations).

\section{Conclusions}
\label{concl}

The coupling of optical power from one OAM mode to different OAM modes due to scintillation is investigated with the aid of the IPE approach, the SPS approach and numerical simulations. We contrast the predictions of the IPE, which is valid under all scintillation conditions, with those of the SPS approach, which is valid only under weak scintillation conditions. It allows us to identify and describe the qualitative behavior of the coupling under various conditions. Using independent MPS numerical simulations, we show that the IPE predictions are favored above the SPS predictions.

\appendix

\section{\label{appenda}Power fraction expressions}

The IPE expressions for the power fraction in the LG modes with $\ell=1,2,3$ when the input LG mode is $\ell=1$ are as follows
\begin{align}
\begin{split}
\eta_{1 \rightarrow 1} = &\left[(u^2+2u+2)v^3+2(u+2)^2u^3v^2\right. \\
& \left.+2(u^2+4u+8)u^6 v\right] \frac{1}{2\Gamma^3} \\
\eta_{1 \rightarrow 2} = &\left[(3u^2+8u+8)v^3+4(u+2)(u+4)u^3v^2\right. \\
& \left.+4(u^2+4u+12)u^6v\right] \frac{\Lambda}{8\Gamma^4} \\
\eta_{1 \rightarrow 3} = &\left[(u^2+3u+3)v^3+(u+2)(u+6)u^3v^2\right. \\
& \left.+(u^2+4u+16)u^6v\right] \frac{\Lambda^2}{4\Gamma^5} ,
\end{split}
\label{123}
\end{align}
where
\begin{align}
\begin{split}
\Gamma & = u^4+4u^3+uv+v \\
\Lambda & = (2u^3+4u^2+v)u \\
v & = 407.7 \mathcal{K}^2 \\
u & = 3.455 \mathcal{W}^{5/3} .
\end{split}
\end{align}
The subscript $1 \rightarrow n$ indicates that the input mode has azimuthal index $\ell=1$ and the overlap mode has azimuthal index $\ell=n$.

In the weak scintillation limit, the expressions simplify to
\begin{align}
\begin{split}
\eta_{1 \rightarrow 1} & = \frac{u^2+2u+2}{2(u+1)^3} \\
\eta_{1 \rightarrow 2} & = \frac{(3u^2+8u+8)u}{8(u+1)^4} \\
\eta_{1 \rightarrow 3} & = \frac{(u^2+3u+3)u^2}{4(u+1)^5} .
\end{split}
\label{123w}
\end{align}
These results are the same as those that can be obtained in direct SPS calculations.

\section*{Acknowledgements}

CMM acknowledges support from the CSIR National Laser Centre.


\end{document}